\documentclass[aps, pra, twocolumn, amsmath, amssymb, superscriptaddress, nofootinbib]{revtex4-1}

\makeatletter
\def\p@subsection{}
\def\p@subsubsection{}
\makeatother

\usepackage[utf8]{inputenc}
\usepackage{graphicx}
\usepackage{units}
\usepackage{color}
\usepackage[pdftex, colorlinks=true, linkcolor=myblue, citecolor=myblue,
urlcolor=myblue]{hyperref}
\usepackage{times}
\usepackage{comment}
\usepackage{graphicx}
\usepackage{amsmath, calc}
\usepackage{mathrsfs}
\usepackage{amsfonts}
\usepackage{amssymb}
\usepackage{bm}
\usepackage{bbm}
\usepackage{color}
\usepackage{array}
\usepackage{units}
\usepackage{tabstackengine}
\stackMath
\usepackage{nicematrix}

\usepackage{empheq} 
\usepackage{cases}

\definecolor{myblue}{rgb}{0,0,1}
\definecolor{myred}{rgb}{1,0,0}

\newcommand{\bra}[1]{\langle #1|}
\newcommand{\ket}[1]{|#1\rangle}

\DeclareMathOperator{\Tr}{Tr}

\begin{document}


\title{Charging a quantum battery from the Bloch sphere}


\author{C. A. Downing}
\email{c.a.downing@exeter.ac.uk} 
\affiliation{Department of Physics and Astronomy, University of Exeter, Exeter EX4 4QL, United Kingdom}

\author{M. S. Ukhtary} 
\affiliation{Research Center for Quantum Physics, National Research and Innovation Agency (BRIN), South Tangerang 15314, Indonesia}


\date{\today}


\begin{abstract}
We reconsider the quantum energetics and quantum thermodynamics of the charging process of a simple, two-component quantum battery model made up of a charger qubit and a single--cell battery qubit. We allow for the initial quantum state of the charger to lie anywhere on the surface of the Bloch sphere, and find the generalized analytical expressions describing the stored energy, ergotropy and capacity of the battery, all of which depend upon the initial Bloch sphere polar angle in a manner evocative of the quantum area theorem. The origin of the ergotropy produced, as well as the genesis of the battery capacity, can be readily traced back to the quantum coherences and population inversions generated (and the balance between these two mechanisms is contingent upon the starting Bloch polar angle). Importantly, the ergotropic charging power and its associated optimal charging time display notable deviations from standard results which disregard thermodynamic considerations. Our theoretical groundwork may be useful for guiding forthcoming experiments in quantum energy science based upon coupled two-level systems.
\end{abstract}


\maketitle



\section{\label{sec_intro}Introduction}

Quantum batteries are quantum mechanical objects tasked with storing energy~\cite{Campaioli2023, Camposeo2025, Ukhtary2024}. In particular, the charging process for quantum batteries has garnered significant theoretical attention, and various protocols have been developed to optimize this energy transfer procedure~\cite{Alicki2013, Vinjanampathy2015, Andolina2018, Farina2019, Ahmadi2023, quadratic2023, Gangwar2024, Ahmadi2024, Shastri2025, Rangkuti2025, Ahmadi2025, Mousavitaha2025}. Meanwhile, a stream of experimental investigations have already started to test the charging abilities of various kinds of quantum batteries built from optical, superconducting and spin architectures~\cite{Quach2022, Joshi2022, Hu2022, Zheng2023, Qu2023, Huang2023, Niu2024}. Thankfully, the latest empirical results suggest reasonable perspectives for building viable quantum energy storage devices in the future~\cite{Hymas2025, Kurman2025}.

Here we are specifically concerned with the quantum nature of the charging process, and in particular how the quantum coherences (off-diagonal elements of the density matrix) of both the energy source and of the battery influence the overall thermodynamic quality of the device. We shall suppose for simplicity that the quantum battery is composed of a charger qubit, which is prepared in some initial quantum state, and a battery qubit which (being completely uncharged at first) starts off in its ground state. When the qubit--qubit coupling is switched on, from some time $t=0$ until some later disconnection time at the instant where $t=T$, energy can be transferred between the charger and the battery subsystems. For all times $t > T$ the battery qubit is in its storage phase. Only a portion of the energy stored in the battery qubit, the so-called ergotropy~\cite{Lenard1978, Pusz1978, Allahverdyan2004}, can be extracted via a cyclic unitary operation. This provides a measure of the quantum thermodynamic quality of the quantum battery, since it describes much work can be reasonably extracted from the device.

\begin{figure}[tb]
 \includegraphics[width=1.0\linewidth]{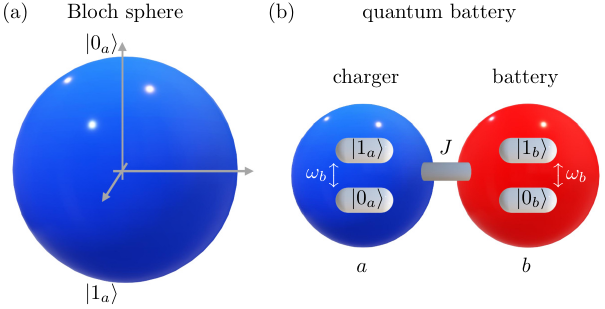}
 \caption{\textbf{Charging a quantum battery from the Bloch sphere.} Panel (a): a geometrical representation of the space of possible quantum states of the qubit modelling the quantum mechanical charger [cf. Eq.~\eqref{eq:zdfdzdfdzf}]. Panel (b): the quantum battery is built from a pair of coupled two-level systems, where the charger is associated with the label $a$ and the battery with $b$. The qubits are each characterized by the transition frequency $\omega_b$ [cf. Eq.~\eqref{eq:dsfsdfsdf}] and they exchange energy at the coupling rate $J$ [cf. Eq.~\eqref{eq:dsfsdfsdffhfhff}] during the charging phase.}
 \label{figCART}
\end{figure}

Let us assume that the composite qubit--qubit system is initially prepared in a quantum state $\ket{\psi}= \ket{\alpha} \otimes \ket{\beta}$, which is constructed from an outer product of the charger and battery subsystem states respectively as follows~\cite{Allen1987, Shore2011}
\begin{align}
\ket{\alpha} &= \cos \left( \tfrac{\theta}{2} \right) \ket{0}_a + \mathrm{e}^{\mathrm{i}\phi} \sin \left( \tfrac{\theta}{2} \right) \ket{1}_a, \label{eq:zdfdzdfdzf} \\
\ket{\beta} &= \ket{0}_b. \label{eq:zdfdzdfdzf222}
\end{align}
The charger qubit (labelled `$a$') is considered to be in an arbitrary superposition state $\ket{\alpha}$ lying on the surface of the Bloch sphere, such that it can be described by the two spherical coordinate parameters $\theta$ and $\phi$ [cf. the cartoon of Fig.~\ref{figCART}~(a)]. The polar angle $\theta$ satisfies the inequality $0 \le \theta \le \pi$ and the azimuthal angle $\phi$ follows $0 \le \phi \le 2\pi$. In particular, three locations of interest on the Bloch sphere are: the north pole where $\theta = 0$ and $\ket{\alpha} = \ket{0}_a$, the south pole where $\theta = \pi$ and essentially $\ket{\alpha} = \ket{1}_a$, and the equator line of latitude where $\theta = \pi/2$ and so $\ket{\alpha} = (\ket{0}_a + \exp{ ( \mathrm{i}\phi ) } \ket{1}_a)/\sqrt{2}$ [cf. Eq.~\eqref{eq:zdfdzdfdzf}]. Clearly, the quantum state $\ket{\alpha} \bra{\alpha}$ has a unitary trace since $\Tr{ \ket{\alpha} \bra{\alpha} } = 1$, and it is a pure quantum state because the purity measure is unity, or $\Tr{( \ket{\alpha} \bra{\alpha} )^2} = 1$. Meanwhile, the battery qubit (tagged `$b$') is assumed to be initially uncharged and so it lies in its ground state [cf. Eq.~\eqref{eq:zdfdzdfdzf222}]. We are interested in finding out how the most general form of a quantum superposition state $\ket{\alpha}$ for the charger qubit affects the charging performance of the combined quantum battery system, both energetically and quantum thermodynamically, and the origins of any nonzero ergotropy created (be it from population inversions or quantum coherences), as well as the role played by the Bloch polar angle $\theta$.

The three-component Hamiltonian operator $\hat{H}$ of the proposed quantum battery system reads [cf. the sketch of Fig.~\ref{figCART}~(b)]
\begin{equation}
\label{eq:sfdfs}
\hat{H} = \hat{H}_a + \hat{H}_b + \hat{H}_{a-b},
\end{equation}
where the energies of the charger qubit and the single--cell battery qubit, as well as the kinetic energy describing the coherent coupling between them (which is switched on during the charging phase $0 < t < T$ only), are captured by the constituent triumvirate of Hamiltonians~\cite{Andolina2018, Farina2019, Shastri2025}
\begin{equation}
\label{eq:dsfsdfsdf}
\hat{H}_a = \omega_b \sigma_a^\dagger \sigma_a, 
\quad\quad
\hat{H}_b = \omega_b \sigma_b^\dagger \sigma_b,
\end{equation}
\begin{equation}
\label{eq:dsfsdfsdffhfhff}
\hat{H}_{a-b} = J \left( \sigma_a^\dagger \sigma_b + \sigma_b^\dagger \sigma_a \right). 
\end{equation}
The pair of qubits both exhibit the common transition frequency $\omega_b$, and they exchange energy at the coupling rate $J$, where the typical experimental coupling regime whereby $J \ll \omega_b$ is assumed to hold. The two flavours $s = \{ a, b\}$ of raising and lower operators appearing in Eq.~\eqref{eq:sfdfs} satisfy the characteristic anticommutation relation $\{ \sigma_s, \sigma_s^\dagger \} = 1$ of two--level systems~\cite{Allen1987, Shore2011}.


\section{\label{sec_mes}Measures of performance}

The capabilities of the proposed quantum battery can be judged from the perspective of both quantum energetics and quantum thermodynamics~\cite{Campaioli2023, Hyperbolic2024}. When considering the energetics, in addition to the total amount of energy $E$ stored in the quantum battery there are three measures of primary interest:
\begin{align}
E \left( t_E \right) &= \max_t{\{ E(t) \}},  \label{eq:sfdsdfsf22} \\
P &= E/t, \label{eq:sfdsdfsf33} \\
P \left( t_P \right) &= \max_t{\{ P(t) \}}. \label{eq:sfdsdfsf44} 
\end{align}
where $t_E$ is the optimal charging time, the instant in time which first corresponds to the maximum stored energy $E(t_E)$. Similarly, the charging power $P$ can also be classified by its optimal charging time $t_P$ and its momentary peak value $P(t_P)$. Furthermore, using Eq.~\eqref{eq:dsfsdfsdf} the inevitable energy fluctuations arising during the charging process can be quantified using the energetic variance~\cite{Friis2018, Sassetti2020}
\begin{align}
\label{eq:dsfdfsf}
\sigma_E^2 &= \langle \hat{H}_b^2 \rangle - \langle \hat{H}_b \rangle^2, \nonumber \\
&= \omega_b^2 \langle \sigma_b^\dagger \sigma_b \rangle \left( 1 - \langle \sigma_b^\dagger \sigma_b \rangle \right). 
\end{align}
Notably when the variance vanishes ($\sigma_E^2=0$), for example when the mean population of the battery qubit is unity, the charging process is entirely noiseless.

Thermodynamically, not all of the energy stored in a quantum system can be used to do work~\cite{Lenard1978, Pusz1978}. The so-called ergotropy $\mathcal{E}$ measures the maximum amount of extractable work compatible with quantum mechanics via the formula~\cite{Allahverdyan2004}
\begin{equation}
\mathcal{E} = E - \textsf{E}, \label{eq:sfdsdfsf55}  
\end{equation}
where the so-called passive state energy $\textsf{E}$, associated with the quantum state from which no further work can be extracted from the system via cyclic processes, has been subtracted off of the stored energy $E$~\cite{Lenard1978, Pusz1978}. Then, in a similar manner to the energetic measures of Eqs.~\eqref{eq:sfdsdfsf22},~\eqref{eq:sfdsdfsf33}~and ~\eqref{eq:sfdsdfsf44}, the following thermodynamic indicators
\begin{align}
\mathcal{E}  \left( t_\mathcal{E}  \right) &= \max_t{\{ \mathcal{E}(t_\mathcal{E}) \}},  \label{eq:sfdsdfsf66} \\
\mathcal{P} &= \mathcal{E}/t, \label{eq:sfdsdfsf77} \\
\mathcal{P}  \left( t_\mathcal{P} \right) &= \max_t{\{ \mathcal{P}  (t_\mathcal{P} ) \}}, \label{eq:sfdsdfsf88} 
\end{align}
define the maximum ergotropy $\mathcal{E} ( t_\mathcal{E} )$ and its charging time $t_\mathcal{E}$, as well as the ergotropic charging power $\mathcal{P}$ and its peak $\mathcal{P}  \left( t_\mathcal{P} \right)$ at some instant $t_\mathcal{P}$. Importantly, these optimal times and maxima will likely deviate from their rather less sophisticated energy-only counterparts listed previously.

The quantum state of the battery qubit subsystem can be described in terms of its first moments and second moments using the $2\times2$ density matrix~\cite{Allen1987, Shore2011}
\begin{equation}
\label{eq:asdasd}
 \rho = \begin{pmatrix}
\langle \sigma_b^\dagger \sigma_b \rangle & \langle \sigma_b \rangle \\
\langle \sigma_b^\dagger  \rangle & 1- \langle \sigma_b^\dagger \sigma_b \rangle 
\end{pmatrix}.
\end{equation}
For some apposite unitary operator $U$, the density matrix $\rho$ can be transformed into its passive state $\varrho = U \rho U^\dagger$, which has the same eigenvalues as $\rho$ but as an ordered diagonal matrix has a lower internal energy, explicitly~\cite{Lenard1978, Pusz1978, Allahverdyan2004}
\begin{equation}
\label{eq:addsddfsdfsdfasd}
\varrho = \begin{pmatrix}
\lambda_- & 0 \\
0 & \lambda_+ 
\end{pmatrix}, 
\end{equation}
\begin{equation}
\label{eq:sdfsdfsdfds}
\lambda_\pm = \tfrac{1 \pm  \sqrt{ 1 + 4 \langle \sigma_b \rangle \langle \sigma_b^\dagger \rangle - 4 \langle \sigma_b^\dagger \sigma_b \rangle \left( 1 - \langle \sigma_b^\dagger \sigma_b \rangle \right) }}{2}. 
\end{equation}
Then the stored energy $E$ and the passive state energy $\textsf{E}$ can be computed directly from Eq.~\eqref{eq:asdasd} and Eq.~\eqref{eq:addsddfsdfsdfasd} as~\cite{Farina2019, Shastri2025}
\begin{align}
\label{eq:dsfxddxfdfsf}
E &= \Tr{ \left(  \hat{H}_b \rho \right) } = \omega_b \langle \sigma_b^\dagger \sigma_b \rangle, \\
\textsf{E} &= \Tr{ \left(  \hat{H}_b \varrho \right) } = \omega_b \lambda_-,  \label{eq:sgdggrghtyy}
\end{align}
which in turn fully defines the ergotropy $\mathcal{E}$ of Eq.~\eqref{eq:sfdsdfsf55} in terms of the readily obtainable first and second moments. Similarly, the so-called antiergotropy $\mathcal{A}$ can be defined as the minimum amount of extractable work congruous with the laws of quantum mechanics~\cite{Yang2023}, such that in analogy to Eq.~\eqref{eq:sfdsdfsf55} it may be found via the formula $\mathcal{A} = E -  \textsf{A}$. Here the active state energy $\textsf{A} = \omega_b \lambda_+$ [cf. Eq.~\eqref{eq:sdfsdfsdfds} for the eigenvalue $\lambda_+$, and Eq.~\eqref{eq:sgdggrghtyy} for the comparable passive state energy $\textsf{E}$]. Then the thermodynamic capacity $\mathcal{K}$ of the quantum battery, which is the amount of work which the battery can transfer during a unitary cycle of evolution, is simply
\begin{equation}
\label{eq:dfgjyytjyjtyjty}
\mathcal{K} = \mathcal{E} - \mathcal{A} = \textsf{A} - \textsf{E} = \omega_b \left( \lambda_+ - \lambda_- \right).
\end{equation}
This important capacity quantity $\mathcal{K}$ captures the ability of the battery to both store and supply energy~\cite{Yang2023}. Overall, this analysis suggests that both quantum coherences and population inversions are likely to be thermodynamically important [cf. Eq.~\eqref{eq:sdfsdfsdfds}], with potential implications for the fundamental explanation of the origin of the ergotropy $\mathcal{E}$ generated in the battery, as well as for the provenance of the battery capacity $\mathcal{K}$ itself. The freedom gained from the tunability of the starting Bloch polar angle $\theta$ should be particularly useful in this regard [cf. Eq.~\eqref{eq:zdfdzdfdzf}]. 
 

\section{\label{sec_dynamics}Quantum dynamics}

\begin{figure*}[tb]
 \includegraphics[width=1.0\linewidth]{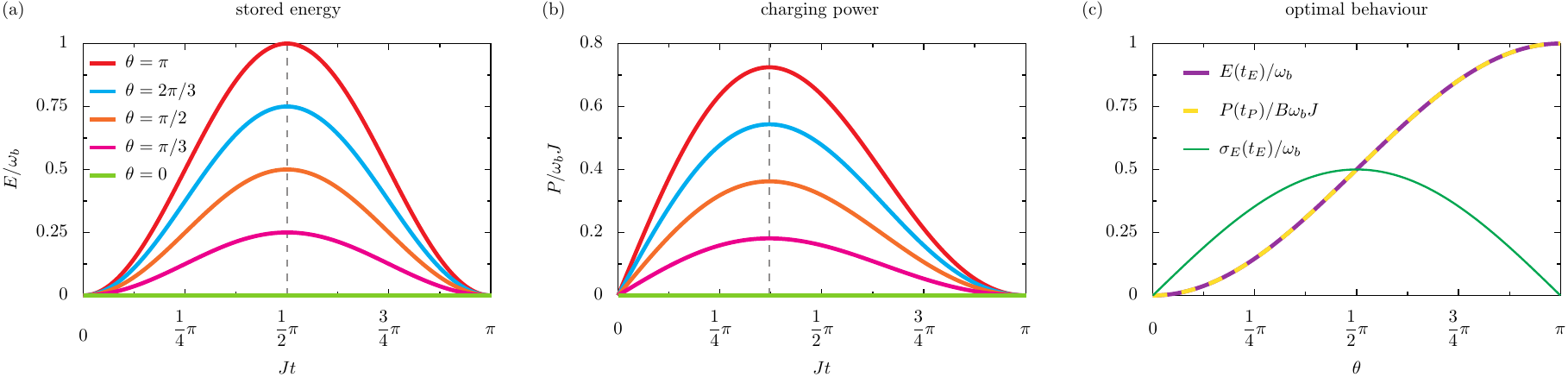}
 \caption{\textbf{Energetics of the quantum battery.} Panel (a): the energy $E$ stored in the quantum battery, in units of the transition frequency $\omega_b$, as a function of the time $t$ which has elapsed (in units of the inverse coupling rate $1/J$) since the coupling was switched on [cf. Eq.~\eqref{eq:sfddddddffsdfsf}]. Dashed line: the optimal time $t_{E}$ [cf. Eq.~\eqref{eq:dvddeddfg}]. Panel (b): the charging power $P$, in units of $\omega_b J$, as a function of time $t$ [cf. Eq.~\eqref{eq:sfgfgfgfgfgdsdfsf}]. Dashed line: the optimal time $t_{P}$ [cf. Eq.~\eqref{eq:dvdxfgxffdvdsgfg}]. Coloured lines in panels (a, b): the various values of the Bloch sphere polar angle $\theta$ considered [cf. Eq.~\eqref{eq:zdfdzdfdzf}]. Panel (c): the optimal stored energy $E(t_{E})$ (purple line) and optimal charging power $P(t_{P})$ (yellow line), in units of $\omega_b$ and $B \omega_b J$ respectively, as a function of the angle $\theta$ [cf. Eq.~\eqref{eq:dfgdgdfgbv} and Eq.~\eqref{eq:dfxgfxgfggdg}]. Thin green line: the energetic standard deviation $\sigma_{E} (t_{E})$ [cf. Eq.~\eqref{eq:sdfsdfssdfsdfsdsf}].}
 \label{FIGenergy}
\end{figure*}

The time evolution of the statistical moments of the quantum battery can be found by applying the trace identity $\Tr{(\rho \mathcal{O} )} = \langle \mathcal{O} \rangle$, which is valid for some given operator $\mathcal{O}$. The Liouville–von Neumann equation $\mathrm{i} \partial_t \rho = [ \hat{H} , \rho ]$ governs how the density matrix $\rho$ of the combined quantum system evolves in time~\cite{Downing2024b}. This theoretical prescription, employed with the full Hamiltonian operator $\hat{H}$ of Eq.~\eqref{eq:sfdfs}, leads to a Schrödinger-like equation for the mean populations of the two qubits~\cite{Farina2019, Shastri2025}
\begin{equation}
\label{eq:dfsdfs}
\mathrm{i} \partial_t \Psi = \mathcal{H} \Psi, 
\end{equation}
where the collection of second moments are contained within $\Psi$, while the dynamical matrix $\mathcal{H}$ accounts for the inter--qubit coupling (at the rate $J$) as follows
\begin{equation}
\label{eq:dfsefsdfsdfdgdfgfsdfs}
\Psi = 
\begin{pmatrix}
\langle \sigma_a^\dagger \sigma_a \rangle  \\
\langle \sigma_b^\dagger \sigma_b \rangle  \\
\langle \sigma_a^\dagger \sigma_b \rangle  \\
\langle \sigma_b^\dagger \sigma_a \rangle  
\end{pmatrix}, 
\quad\quad
\mathcal{H} = 
\begin{pmatrix}
0 & 0 & J & -J \\
0 & 0 & -J & J\\
J & -J & 0 & 0 \\
-J & J & 0 & 0 
\end{pmatrix}. 
\end{equation}
The exact solution to this system of equations, subject to the initial conditions at $t=0$ as implied by Eq.~\eqref{eq:zdfdzdfdzf} and Eq.~\eqref{eq:zdfdzdfdzf222}, reads
\begin{align}
\langle \sigma_a^\dagger \sigma_a \rangle  &= \sin^2 \left( \tfrac{\theta}{2} \right) \cos^2 \left( J t \right), \label{eq:sfdxdfxdfdsfsdfsf} \\
\langle \sigma_b^\dagger \sigma_b \rangle &= \sin^2 \left( \tfrac{\theta}{2} \right) \sin^2 \left( J t \right),  \label{eq:dsffsvsf} \\
\langle \sigma_a^\dagger \sigma_b \rangle &= \langle \sigma_b^\dagger \sigma_a \rangle ^\ast = -\tfrac{\mathrm{i}}{2} \sin^2 \left( \tfrac{\theta}{2} \right) \sin \left( 2 J t \right), \label{eq:sfddsfgvb} 
\end{align}
which are necessary for calculating the energy $E$ stored in the quantum battery following Eq.~\eqref{eq:dsfxddxfdfsf}. Alongside the conventional Rabi oscillations present in these elementary expressions, the trigonometric prefactor recalls the celebrated quantum area theorem for a single two-level system driven by an optical pulse, where the integrated pulse area plays the role of the Bloch polar angle $\theta$~\cite{Allen1987, Shore2011}. Necessarily, the mean populations of Eq.~\eqref{eq:sfdxdfxdfdsfsdfsf} and Eq.~\eqref{eq:dsffsvsf} satisfy the time-independent identity $\langle \sigma_a^\dagger \sigma_a \rangle + \langle \sigma_b^\dagger \sigma_b \rangle = \sin^2 \left( \theta/2 \right)$, since the full Hamiltonian operator $\hat{H}$ of Eq.~\eqref{eq:sfdfs} conserves the total excitation number throughout the charging process. Equivalently, the relevant commutator $[\hat{N}, \hat{H}] = 0$, where $\hat{N} = \sigma_a^\dagger \sigma_a + \sigma_b^\dagger \sigma_b$ is the pertinent number operator.

Similar to Eq.~\eqref{eq:dfsdfs}, the equation of motion for the first moments, which are important for understanding the ergotropy $\mathcal{E}$ since the quantity $\langle  \sigma_b \rangle$ enters the passive state energy $\textsf{E}$ definition of Eq.~\eqref{eq:sgdggrghtyy} via the eigenvalue $\lambda_-$, is given by~\cite{Farina2019, Shastri2025}
\begin{equation}
\label{eq:sdfsfdsdfdsf}
\mathrm{i} \partial_t \psi = \mathcal{M} \psi. 
\end{equation}
The correlators in $\psi$ and the $4 \times 4$ dynamical matrix $\mathcal{M}$ are
\begin{equation}
\label{eq:dfsefsdfsdfsdfs}
\psi = 
\begin{pmatrix}
\langle  \sigma_a \rangle  \\
\langle  \sigma_b \rangle  \\
\langle \sigma_a^\dagger \sigma_a \sigma_b \rangle  \\
\langle \sigma_b^\dagger \sigma_b \sigma_a \rangle  
\end{pmatrix}, 
\quad
\mathcal{M} = 
\begin{pmatrix}
\omega_0 & J & -2J & 0 \\
J & \omega_0 & 0 & -2J\\
0 & 0 & \omega_0 & -J \\
0 & 0 & -J & \omega_0 
\end{pmatrix}, 
\end{equation}
such that the exact and analytical solution to Eq.~\eqref{eq:sdfsfdsdfdsf} is simply
\begin{align}
\langle \sigma_a \rangle  &= \tfrac{1}{2} \mathrm{e}^{\mathrm{i}\phi} \sin \left( \theta \right) \cos \left( J t \right) \mathrm{e}^{-\mathrm{i}\omega_b t}, \label{eq:fsvbb} \\
\langle  \sigma_b \rangle &= -\tfrac{\mathrm{i}}{2} \mathrm{e}^{\mathrm{i}\phi} \sin \left( \theta \right) \sin \left( J t \right) \mathrm{e}^{-\mathrm{i}\omega_b t},  \label{eq:thtyhr} \\
\langle \sigma_a^\dagger \sigma_a \sigma_b \rangle &= \langle \sigma_b^\dagger \sigma_b \sigma_a \rangle = 0. \label{eq:fsevcrds} 
\end{align}
The knowledge of the quantum moments derived here is utilized in what follows, firstly in an energetic analysis and then in a more sophisticated quantum thermodynamic treatment.


\section{\label{sec_model}Energetics}

The dynamical energy $E$ stored in the quantum battery follows from the formal definition of Eq.~\eqref{eq:dsfxddxfdfsf} and the mean population of Eq.~\eqref{eq:dsffsvsf} as
\begin{equation}
\label{eq:sfddddddffsdfsf} 
E_\theta = \omega_b \sin^2 \left( \tfrac{\theta}{2} \right) \sin^2 \left( J t \right),
\end{equation}
which clearly exhibits a squared sinusoidal dependence on the polar angle $\theta$, which itself arises from the nature of the geoposition of the initial quantum state of the charger qubit lying on the surface of the Bloch sphere [cf. Eq.~\eqref{eq:zdfdzdfdzf}]. The expression of Eq.~\eqref{eq:sfddddddffsdfsf} then evokes the physics of the quantum area theorem~\cite{Allen1987, Shore2011}. The stored energy $E$ is plotted in Fig.~\ref{FIGenergy}~(a) as a function of the time which has elapsed since the coupling was switched on for several values of the Bloch polar angle $\theta$ (coloured lines). Clearly, it is energetically advantageous to have as large an initial population in the charger as possible, with the best case being $\theta = \pi$ (south pole) where the prepared charger state $\ket{\alpha} = \ket{1}_a$ (red line)~\cite{Andolina2018}. In the extreme case of $\theta = 0$ (north pole), corresponding to a trivial initial charger state $\ket{\alpha} = \ket{0}_a$, there are no excitations in the system from the moment of preparation onwards (green line). Notably, the peak in energy $E$ in Fig.~\ref{FIGenergy}~(a) is reached at a universal time which is independent of the Bloch polar angle $\theta$, leading to the optimal charging time $t_{E}$ and the maximum stored energy $E(t_{E})$ [cf. Eq.~\eqref{eq:sfdsdfsf22}]
\begin{align}
t_{E} &= \tfrac{\pi}{2 J}, \label{eq:dvddeddfg} \\
E_\theta \left( t_{E} \right) &= \omega_b \sin^2 \left( \tfrac{\theta}{2} \right),  \label{eq:dfgdgdfgbv} 
\end{align}
as marked by the dashed vertical grey line in Fig.~\ref{FIGenergy}~(a). This brief analysis suggests the moment at which the charger should ideally be disconnected from the battery is insensitive as to how the charger qubit was initially prepared on the surface of the Bloch sphere.

The associated charging power $P$ follows directly from Eq.~\eqref{eq:sfdsdfsf33} with Eq.~\eqref{eq:sfddddddffsdfsf} as
\begin{equation}
\label{eq:sfgfgfgfgfgdsdfsf} 
P_\theta = \omega_b \sin^2 \left( \tfrac{\theta}{2} \right) \tfrac{\sin^2 \left( J t \right)}{t},
\end{equation}
which is plotted in Fig.~\ref{FIGenergy}~(b), both as a function of the charging time and for several values of the Bloch polar angle $\theta$ (coloured lines) which correspond to the cases already studied in panel~(a). Unlike the stored energy $E$, these power curves are not periodic in time due to the temporal inverse-linear decay inherent to Eq.~\eqref{eq:sfgfgfgfgfgdsdfsf}. Notably, the time of maximum power $t_{P}$ and the peak power $P (t_{P})$ are given by the rather neat expressions [cf. Eq.~\eqref{eq:sfdsdfsf44}]
\begin{align}
t_{P} &= \tfrac{A}{J},
&& A = 1.165\ldots, \label{eq:dvdxfgxffdvdsgfg} \\
P_\theta \left( t_{P} \right) &= B \omega_b J \sin^2 \left( \tfrac{\theta}{2} \right), 
&& B = 0.724\ldots, \label{eq:dfxgfxgfggdg} 
\end{align}
where the number $A$ is a solution to the transcendental trigonometric equation $\tan (A) = 2 A$, while the second number $B$ arises from evaluating the expression $B = \sin^2 (A)/A$. As was the situation for the stored energy $E$, the charging time $t_{P}$ (dashed vertical grey line) is independent of the Bloch polar angle $\theta$, but the special instant in time $t_{P}$ is always arrived at faster than $t_{E}$ due to the temporal decay innate to Eq.~\eqref{eq:sfgfgfgfgfgdsdfsf}.

The optimal energy $E_\theta(t_{E})$ (purple line) and the optimal charging power $P_\theta(t_{P})$ (yellow line) are plainly functionally identical, as displayed in Fig.~\ref{FIGenergy}~(c) as a function of the Bloch polar angle $\theta$ [cf. Eq.~\eqref{eq:dfgdgdfgbv} and Eq.~\eqref{eq:dfxgfxgfggdg}]. This simple plot confirms the intuitive understanding that one should start from the south pole of the Bloch sphere ($\theta = \pi$) in order to maximise the energetic performance of the quantum battery. Meanwhile, the energy fluctuations arising during the charging process can be characterised by the energy variance $\sigma_{E}^2$ of Eq.~\eqref{eq:dsfdfsf} as
\begin{equation}
\label{eq:sdfsdfsdsf}
\sigma_{E}^2 = \omega_b^2 \sin^2 \left( \tfrac{\theta}{2} \right) \sin^2 \left( J t \right) \left( 1- \sin^2 \left( \tfrac{\theta}{2} \right) \sin^2 \left( J t \right)\right), 
\end{equation}
which is notably nonzero for any nontrivial initial charger state where the Bloch polar angle $\theta \ne 0$. Specifically at the optimal charging time $t_E$ of Eq.~\eqref{eq:dvddeddfg}, the trigonometric expression of Eq.~\eqref{eq:sdfsdfsdsf} collapses into the energetic standard deviation
\begin{equation}
\label{eq:sdfsdfssdfsdfsdsf}
\sigma_{E} (t_{E}) = \tfrac{\omega_b}{2} \sin \left( \theta \right),
\end{equation}
as marked by the thin green line in Fig.~\ref{FIGenergy}~(c). Hence when the maximum stored energy of $\omega_b$ is achieved, with the Bloch polar angle $\theta = \pi$, there are pleasingly no energetic fluctuations. However, on the Bloch equator with $\theta = \pi/2$ the maximum energetic standard deviation of $\omega_b/2$ is unfortunately realized. Importantly, these elementary energetic results are liable to change when the quantum thermodynamic nature of the quantum battery is properly considered, as can now be discussed.


\section{\label{sec_therm}Thermodynamics}

\begin{figure*}[tb]
 \includegraphics[width=1.0\linewidth]{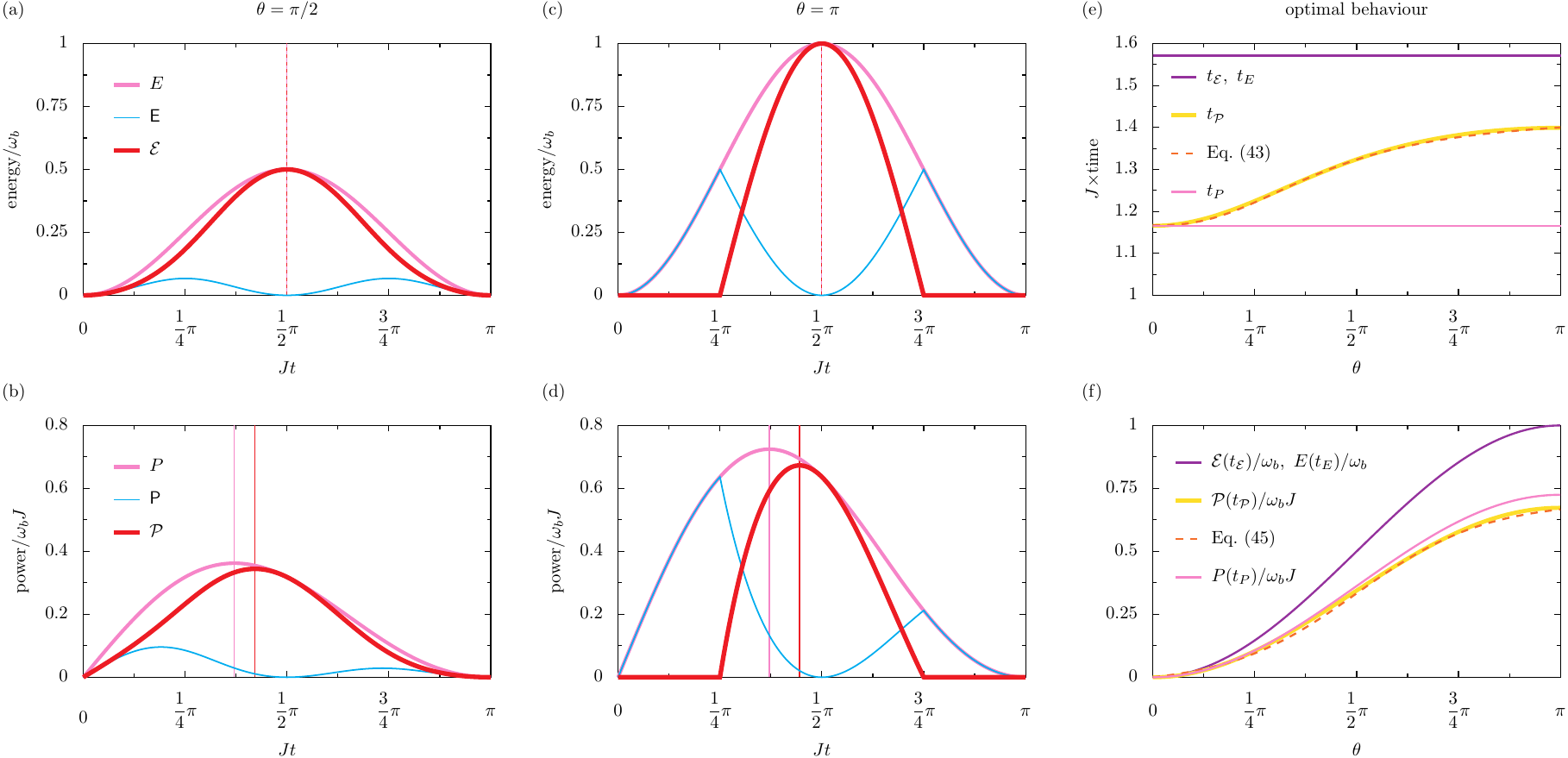}
 \caption{\textbf{Thermodynamics of the quantum battery.} Panel (a): the energy $E$ (pink line), passive state energy $\textsf{E}$ (cyan line), and ergotropy $\mathcal{E}$ (red line) in the quantum battery, all in units of the transition frequency $\omega_b$, as a function of the time $t$ which has elapsed (in units of the inverse coupling rate $1/J$) since the coupling was switched on [cf. Eq.~\eqref{eq:sfddddddffsdfsf} and Eq.~\eqref{eq:fsdfs}]. Vertical pink and red lines: the optimal times $t_{E}$ and $t_{\mathcal{E}}$ respectively [cf. Eq.~\eqref{eq:dvddeddfg} and Eq.~\eqref{eq:dvdfg}].  Panel (b): the charging power $P$ (pink line), passive state charging power $\textsf{P}$ (cyan line), and ergotropic charging power $\mathcal{P}$ (red line), all in units of $\omega_b J$, as a function of time $t$ [cf. Eq.~\eqref{eq:sfgfgfgfgfgdsdfsf}]. Vertical pink and red lines: the optimal times $t_{P}$ and $t_{\mathcal{P}}$ respectively [cf. Eq.~\eqref{eq:dvdxfgxffdvdsgfg} and Eq.~\eqref{eq:dgsfsdf}]. In this column, the Bloch sphere polar angle $\theta = \pi/2$. Panels (c, d): as for the first column, but where $\theta = \pi$. Panel (e): the optimal charging times for energy and ergotropy $t_{E}$ and $t_{\mathcal{E}}$ (purple line), for charging power $t_P$ (pink line) and for ergotropic charging power $t_{\mathcal{P}}$ (yellow line), all in units of $1/J$, as a function of $\theta$ [cf. Eq.~\eqref{eq:dvddeddfg}, Eq.~\eqref{eq:dvdxfgxffdvdsgfg} and Eq.~\eqref{eq:dvdfg}]. Dashed orange line: the analytic approximation of Eq.~\eqref{eq:dgsfsdf}. Panel (f): the optimal quantities associated with the charging times of panel (e), including the peak energy $E(t_{E})$ and peak ergotropy $\mathcal{E} (t_{\mathcal{E}})$ (purple line), both in units of $\omega_b$, and the peak ergotropic charging power $\mathcal{P} (t_{\mathcal{P}})$ (yellow line) and peak charging power $P(t_{P})$ (pink line), both in units of $\omega_b J$, all as a function of $\theta$ [cf. Eq.~\eqref{eq:dfgdgdfgbv}, Eq.~\eqref{eq:dfxgfxgfggdg} and Eq.~\eqref{eq:dfgdg}]. Dashed orange line: the analytic approximation of Eq.~\eqref{eq:sdfsdf}.}
 \label{FIGthermo}
\end{figure*}

The ergotropy $\mathcal{E}$ achieved by the quantum battery can be computed from the definition in Eq.~\eqref{eq:sfdsdfsf55} along with the stored energy $E$ from Eq.~\eqref{eq:sfddddddffsdfsf} and the passive state energy $\textsf{E}$, which is defined in Eq.~\eqref{eq:sgdggrghtyy} [via the eigenvalue of Eq.~\eqref{eq:sdfsdfsdfds}] and made explicit by the first moment of Eq.~\eqref{eq:thtyhr}. The analytical expression for the ergotropy $\mathcal{E}$, for arbitrary values of the Bloch sphere polar angle $\theta$, as well as for the limiting case of $\theta = \pi$ immediately below it, are then given by
\begin{widetext}
\begin{equation}
\label{eq:fsdfs}
\mathcal{E}_\theta = \omega_b \left[ \sin^2 \left( \tfrac{\theta}{2} \right) \sin^2 \left( J t \right) - \tfrac{1}{2} \left( 1 - \sqrt{1 - \sin^4 \left( \tfrac{\theta}{2} \right) \sin^2 \left( 2 J t \right)} \right) \right],
\end{equation}
\begin{align}
\label{eq:fdgfgdfg}
\mathcal{E}_\pi &= \omega_b \sin^2 \left( J t \right) - \tfrac{\omega_b}{2} \left( 1 - | \cos \left( 2 J t\right) | \right), \nonumber \\
 &= \begin{cases}
  0,  & 0 \le J t \le \tfrac{\pi}{4}  ~\text{and}~ \pi \left( m + \tfrac{3}{4} \right) \le J t \le \pi \left( m + \tfrac{5}{4} \right), \\
 - \omega_b \cos \left( 2 J t \right),  & \pi \left( m + \tfrac{1}{4} \right) < J t < \pi \left( m + \tfrac{3}{4} \right),
\end{cases}
\end{align}
\end{widetext}
where $m = 0, 1, 2, \ldots$ is a non-negative integer. The specific result of Eq.~\eqref{eq:fdgfgdfg} for $\mathcal{E}_\pi$ recovers the standard result for a conventional qubit battery, as is also quoted in the Supplemental Material of Ref~\cite{Carrasco2022} for example. Meanwhile Eq.~\eqref{eq:fsdfs} provides the pleasant generalization to the case of charging from an initial position anywhere on the surface of the Bloch sphere (as governed by the polar angle $\theta$ solely, the azimuthal angle $\phi$ is irrelevant here due to the symmetry of the problem).

The general and dynamical ergotropy $\mathcal{E}$ result of Eq.~\eqref{eq:fsdfs} is displayed by the thick red lines in Fig.~\ref{FIGthermo}~(a)~and~(c) for the example cases of the Bloch polar angle $\theta = \pi/2$ and $\theta = \pi$ respectively. The corresponding stored energies $E$ (thin pink lines) and passive state energies $\textsf{E}$ (thin cyan lines) are also shown for completeness. Notably, in general the ergotropy $\mathcal{E}$ is lower than the stored energy $E$ because a portion of the energy remains in the passive state [cf. Eq.~\eqref{eq:sfdsdfsf55}]. However, at the specific instant of maximum ergotropy (thin pink vertical lines), where the optimal behaviour is described by [cf. Eq.~\eqref{eq:sfdsdfsf66}]
\begin{align}
t_{\mathcal{E}} &= \tfrac{\pi}{2 J}, \label{eq:dvdfg} \\
\mathcal{E}_\theta \left( t_{\mathcal{E}} \right) &= \omega_b \sin^2 \left( \tfrac{\theta}{2} \right),  \label{eq:dfgdg} 
\end{align}
the ergotropic and energetic values coincide exactly [cf. Eq.~\eqref{eq:dvddeddfg} and Eq.~\eqref{eq:dfgdgdfgbv}], which is a notable advantage of this kind of two-component qubit quantum battery. There are striking exact ergotropic zeroes in Fig.~\ref{FIGthermo}~(c) for some chunks of time, exactly as predicted by Eq.~\eqref{eq:fdgfgdfg}, due to the periodic lack of any population inversion and zero coherences when the Bloch polar angle $\theta = \pi$. This fact follows directly from the moments defined earlier on: the mean population of Eq.~\eqref{eq:dsffsvsf} cannot rise above one half during certain intervals of time preventing population inversion, and the quantum coherence of Eq.~\eqref{eq:thtyhr} vanishes identically. Elsewhere, for example with $\theta = \pi/2$ in Fig.~\ref{FIGthermo}~(a), there are no finite periods of time with exact ergotropic zeroes because at least one mechanism, either population inversion of quantum coherence, is nonzero -- this directly leads to a passive state energy $\textsf{E}$ which is different from the stored energy $E$.

\begin{figure*}[tb]
 \includegraphics[width=1.0\linewidth]{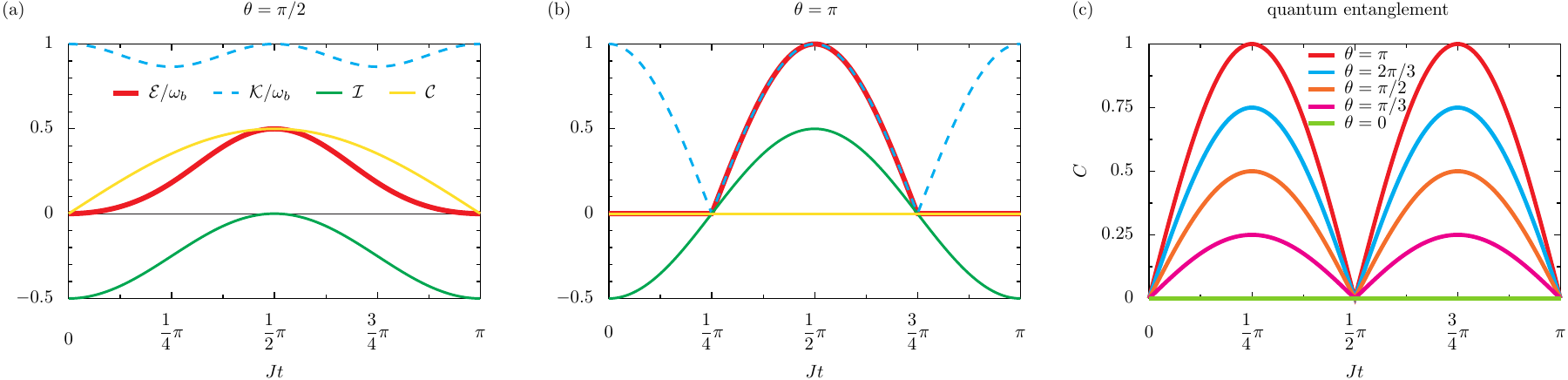}
 \caption{\textbf{The origin of the ergotropy and the genesis of the battery capacity.} Panel (a): the ergotropy $\mathcal{E}$ (red line) and capacity $\mathcal{K}$ (dashed cyan line), both in units of the transition frequency $\omega_b$, as a function of the time $t$ which has elapsed (in units of the inverse coupling rate $1/J$) since the coupling was switched on [cf. Eq.~\eqref{eq:dgffgfg} and Eq.~\eqref{eq:dgffgfg22222}]. Green line: the population inversion parameter $\mathcal{I}$ [cf. Eq.~\eqref{eq:sfdsgdgfdfg}]. Yellow line: the coherence parameter $\mathcal{C}$ [cf. Eq.~\eqref{eq:sfdsgdgfdfg222}]. In this panel the Bloch sphere polar angle $\theta = \pi/2$. Panel (b): as for panel~(a) but where $\theta = \pi$. Panel (c): the dynamical concurrence $C$ generated in the two--qubit system [cf. Eq.~\eqref{eq:conccc}]. Coloured lines: the various values of the Bloch sphere polar angle $\theta$ considered [cf. Eq.~\eqref{eq:zdfdzdfdzf}].}
 \label{FIGentangle}
\end{figure*}

Similar to the energetic plots in the upper panels of Fig.~\ref{FIGthermo}, the associated charging power $P$ (pink lines), passive state charging power $\textsf{P} = \textsf{E} /t$ (cyan lines) and ergotropic charging power $\mathcal{P}$ (red lines) are plotted in the lower panels~(b)~and~(d) of Fig.~\ref{FIGthermo}, for the cases corresponding to upper panels~(a)~and~(c) respectively [cf. Eq.~\eqref{eq:sfdsdfsf77}]. Most importantly, the optimal charging times for power (pink vertical lines) and ergotropic power (red vertical lines) are now clearly distinct from each other, highlighting the importance of a proper thermodynamic study of this quantum battery system in order to accurately gauge its charging performance.

The optimal behaviour of the quantum battery is analysed more generally in the final column of Fig.~\ref{FIGthermo} [cf. Eq.~\eqref{eq:sfdsdfsf66} and Eq.~\eqref{eq:sfdsdfsf88}]. The best charging time $t_{\mathcal{P}}$, that which maximises the vital ergotropic charging power $\mathcal{P}$, is displayed with the thick yellow line in Fig.~\ref{FIGthermo}~(e) as a function of the Bloch sphere polar angle $\theta$. Clearly, unlike the corresponding times for ergotropy and energy (the purple line is for both) [cf. Eq.~\eqref{eq:dvddeddfg} and Eq.~\eqref{eq:dvdfg}] and for normal charging power (pink line) [cf. Eq.~\eqref{eq:dvdxfgxffdvdsgfg}], there is a significant initial condition dependence through the Bloch polar angle $\theta$. The two limiting cases where $\theta = 0$ and $\theta = \pi$ can be analysed exactly, leading to the twin results for $t_{\mathcal{P}_\theta}$ as follows
\begin{equation}
t_{\mathcal{P}_0} = \tfrac{A}{J},
\quad\quad
t_{\mathcal{P}_\pi} = \tfrac{C}{J},
\quad\quad
 C = 1.399\ldots, \label{eq:ddsfdfsvdxfgxfgfg} 
\end{equation}
where $A$ was already defined in Eq.~\eqref{eq:dvdxfgxffdvdsgfg}, and where the number $C$ solves the transcendental equation $1+2C\tan(2C) = 0$. The interesting behaviour shown in Fig.~\ref{FIGthermo}~(e) for the ergotropic charging power can be analytically captured within the spirit of approximation theory, where the asymmetric sigmoid functional shape of the curve suggests the following functional approximation (dashed orange line)
\begin{equation}
\label{eq:dgsfsdf}
t_{\mathcal{P}_\theta}  \simeq \frac{7}{5} \frac{ \frac{5}{6} + \left( \frac{1}{6} + \frac{9\sqrt{3}}{32} \pi^{\frac{5}{2}} \right) \left( \frac{\theta}{\pi} \right)^{\frac{5}{2}}  }{1 + \frac{9\sqrt{3}}{32} \theta^{\frac{5}{2}}} \frac{1}{J},
\end{equation}
which (in units of $1/J$) can be seen to exhibit the pleasing values $7/6 = 1.16\ldots$ and $7/5 = 1.4$ at the two extrema where $\theta = 0$ and $\theta = \pi$ respectively, approximating the exact results of Eq.~\eqref{eq:ddsfdfsvdxfgxfgfg} rather well. Overall, the disparate temporal results shown in Fig.~\ref{FIGthermo}~(e) reveal that it is essential to have a proper understanding of the ergotropic charging power when judging the optimal disconnection time between the charger and the battery qubit components -- energetics alone is not enough.

The corresponding optimal ergotropic charging power $\mathcal{P}(t_{\mathcal{P}})$ is similarly plotted in the lower panel of Fig.~\ref{FIGthermo}~(f) as the thick yellow line, along with the optimal energy $E(t_{E})$ and ergotropy $\mathcal{E} (t_{\mathcal{E}})$ (the purple line is for both) [cf. Eq.~\eqref{eq:dfgdgdfgbv} and Eq.~\eqref{eq:dfgdg}], and the optimal charging power $P(t_{P})$ (pink line) [cf. Eq.~\eqref{eq:dfxgfxgfggdg}]. Most notably, the optimal charging power overestimates the proper and ergotropic charging power for all nontrivial initial charger states, as characterized by a nonzero polar angle $\theta$. For example, the maximum ergotropic charging power when $\theta = \pi$ is
\begin{equation}
\mathcal{P}_\pi \left( t_{P_\pi} \right) = D \omega_b J, 
\quad\quad\quad
 D = 0.673\ldots, \label{eq:dfxgfssdfsdfdfsdfxgfggdg} 
\end{equation}
where the numerical prefactor $D = - \cos (2C)/C$ follows directly from Eq.~\eqref{eq:ddsfdfsvdxfgxfgfg}. This clearly underperforms the normal charging power peak of $P_\pi ( t_{P} ) = B \omega_b J$, as follows straight from Eq.~\eqref{eq:dfxgfxgfggdg}. More pleasantly (from a mathematical point of view), the symmetric sigmoid-like curve of $\mathcal{P} (t_{\mathcal{P}})$, as displayed in Fig.~\ref{FIGthermo}~(f) by the thick yellow line, resembles the hyperbolic tangent analytic approximation (dashed orange line)
\begin{equation}
\label{eq:sdfsdf}
\mathcal{P}_\theta \left( t_{\mathcal{P}_\theta} \right) \simeq \frac{\omega_b J}{3} \left( \frac{ \tanh \left( \theta - \frac{\pi}{2} \right) }{ \tanh \left( \frac{\pi}{2} \right) } + 1 \right).
\end{equation}
This expression (in units of $\omega_b J$) can be seen to exhibit the characteristic values $0$, $1/3$ and $2/3$ at the key Bloch polar angle values of $\theta = 0$, $\pi/2$ and $\pi$ respectively, closely mimicking the limiting case of Eq.~\eqref{eq:dfxgfssdfsdfdfsdfxgfggdg} for example.


\section{\label{sec_origin}Underlying mechanisms}

The interplay between population inversion and quantum coherences for the resulting performance of the quantum battery (for example, as judged by the ergotropy and battery capacity) can be seen more clearly by introducing the population inversion parameter $\mathcal{I}$ and the coherence parameter $\mathcal{C}$ via
\begin{equation}
\label{eq:fgbdgfnhjjhfjgh}  
\mathcal{I} = \langle \sigma_b^\dagger \sigma_b \rangle -\tfrac{1}{2},
\quad\quad\quad
\mathcal{C} = | \langle \sigma_b \rangle |.
\end{equation}
The stored energy $E$, passive state energy $\textsf{E}$ and active state energy $\textsf{A}$ then become [cf. Eq.~\eqref{eq:sdfsdfsdfds}, Eq.~\eqref{eq:dsfxddxfdfsf} and Eq.~\eqref{eq:sgdggrghtyy}]
\begin{align}
E &= \omega_b \left( \tfrac{1}{2} + \mathcal{I} \right),  \label{eq:dgffgfg33333} \\
\textsf{E} &= \omega_b \left( \tfrac{1}{2} - \sqrt{ \mathcal{I}^2 + \mathcal{C}^2 } \right), \label{eq:dgffgfg44444}  \\
\textsf{A} &= \omega_b \left( \tfrac{1}{2} + \sqrt{ \mathcal{I}^2 + \mathcal{C}^2 } \right). \label{eq:dgffgfg55555}   
\end{align}
The useful decomposition of Eq.~\eqref{eq:fgbdgfnhjjhfjgh} also leads directly to the general formulae for the qubit ergotropy $\mathcal{E}$, antiergotropy $\mathcal{A}$ and battery capacity $\mathcal{K}$ as
\begin{align}
\mathcal{E} &= \omega_b \left( \mathcal{I} + \sqrt{ \mathcal{I}^2 + \mathcal{C}^2 } \right), \label{eq:dgffgfg} \\
\mathcal{A} &= \omega_b \left( \mathcal{I} - \sqrt{ \mathcal{I}^2 + \mathcal{C}^2 } \right), \label{eq:dgffgfgYYYYY} \\
\mathcal{K} &= 2\omega_b  \sqrt{ \mathcal{I}^2 + \mathcal{C}^2 }. \label{eq:dgffgfg22222}  
\end{align}
Notably, in the limiting case where there are no coherences ($\mathcal{C} = 0$), which occurs for all charging times when the Bloch polar angle $\theta = \pi$ in the model presented here [cf. Eq.~\eqref{eq:thtyhr}], it is essential to have a population inversion ($\mathcal{I} > 0$) in order to generate nonzero ergotropy since Eq.~\eqref{eq:dgffgfg} collapses into
\begin{equation}
\label{eq:dgffgdfdffg}  
\mathcal{E} = \begin{cases}
  2 \omega_b \mathcal{I},  & \mathcal{I} > 0, \\
  0, & \mathcal{I} \le 0,
\end{cases} 
\quad\quad\quad
 \left( \mathcal{C} = 0 \right).
\end{equation}
During the periods of time whereby the battery qubit is in a pure population inversion state the ergotropy is then equal to the capacity, or $\mathcal{E} = \mathcal{K}$ directly from Eq.~\eqref{eq:dgffgfg22222}. This is because the active state has been achieved, which also means that the antiergotropy vanishes, or simply $\mathcal{A}=0$ from Eq.~\eqref{eq:dgffgfgYYYYY}. Alternatively, for the case when there is a balanced population between the ground and excited state of the battery qubit ($\mathcal{I} = 0$), a situation corresponding for example to the angle $\theta = \pi/2$ and the specific charging time $t = \pi/2J$ in the considered model [cf. Eq.~\eqref{eq:dsffsvsf}], only nonzero quantum coherences contribute to the ergotropy following the formula
\begin{equation}
\label{eq:dgffgfgfgfgfggdfdffg}  
\mathcal{E} = \omega_b  \mathcal{C}, \quad\quad\quad
 \left( \mathcal{I} = 0 \right).
\end{equation}
Furthermore, from Eq.~\eqref{eq:dgffgfg22222} only half of the overall battery capacity $\mathcal{K} = 2 \omega_b  \mathcal{C}$ has been achieved. Both of these two scenarios are represented graphically in Fig.~\ref{FIGentangle}~(a, b), where the dynamical ergotropy $\mathcal{E}$ (red lines) is displayed alongside the population inversion parameter $\mathcal{I}$ (green lines) and the coherence parameter $\mathcal{C}$ (yellow lines), as well as the battery capacity $\mathcal{K}$ (dashed cyan lines). Away from the two extremes cases of Eq.~\eqref{eq:dgffgdfdffg} and Eq.~\eqref{eq:dgffgfgfgfgfggdfdffg}, both population inversion and quantum coherences generally contribute towards the overall ergotropy $\mathcal{E}$ following the exact formula of Eq.~\eqref{eq:dgffgfg}, which with the explicit forms
\begin{align}
\label{eq:sfdsgdgfdfg}  
\mathcal{I}_\theta &= \sin^2 \left( \tfrac{\theta}{2} \right) \sin^2 (Jt) - \tfrac{1}{2}, \\
\mathcal{C}_\theta &= \tfrac{1}{2}\sin (\theta) |\sin(Jt)|, \label{eq:sfdsgdgfdfg222}  
\end{align}
reproduces the exact expression of Eq.~\eqref{eq:fsdfs}. This decompositional analysis highlights the utility of the model considered in this work for explaining the ultimate origin of qubit ergotropy. Similarly, the particular expression for the battery capacity $\mathcal{K}$ follows directly from the general formula of Eq.~\eqref{eq:dgffgfg22222} as
\begin{equation}
\label{eq:fghjhgkuguk}  
\mathcal{K}_\theta = \omega_b \sqrt{ 1 - \sin^4 \left( \tfrac{\theta}{2} \right) \sin^2 \left( 2 J t \right) }, 
\end{equation}
which is plotted as the dashed cyan lines in Fig.~\ref{FIGentangle}~(a, b). Clearly, the battery capacity is maximal (for any value of $\theta$) when the dimensionless time $J t = m\pi/2$ ($m$ being a non-negative integer), but the fundamental reason for this capacity maximization can be quite different. For the subset of times where $J t = m\pi$ the coherences and the mean population of the battery both vanish (that is, $\mathcal{C}_\theta = 0$ and $\mathcal{I}_\theta = -1/2$), such that the qubit has the full ability to store the maximum amount of energy, or $\mathcal{K}_\theta = \omega_b$. Conversely, for the cases when $J t = (m+1/2)\pi$, the coherence parameter $\mathcal{C}_\theta = \sin (\theta)/2$ and the population inversion parameter $\mathcal{I}_\theta = -\cos (\theta)/2$, such that both of these pathways to the ultimate battery capacity $\mathcal{K}_\theta = \omega_b$ contribute in harmony. This interplay may also be deduced by looking at the green and yellow curves of Fig.~\ref{FIGentangle}~(a, b), especially at the instant $J t = \pi/2$.

Taken together, the exact ergotropy and battery capacity expressions provided [cf. Eq.~\eqref{eq:fsdfs} and Eq.~\eqref{eq:fghjhgkuguk}], both of which are valid for an arbitrary initial charger state lying anywhere on the surface of the Bloch sphere, provide the groundwork analysis of the proposed qubit--qubit quantum battery. Significant insight into the genesis of these key thermodynamic quantities has been gained by both an appropriate decomposition and the adjustability of the influential angle $\theta$.


\section{\label{sec_ent}Entanglement}

Some recent studies have noted the importance of quantum entanglement for the inner workings of some kinds of quantum batteries~\cite{Zhang2025, Bai2025}. To investigate the links between entanglement and performance for the quantum battery proposed in Eq.~\eqref{eq:sfdfs}, we first solve the first-order differential equation formed from the quantum master equation discussed at the start of Sec.~\ref{sec_dynamics}, that is $\partial_t \rho = \mathcal{L} \rho$, where $\rho$ is now a sixteen-dimensional density vector (coming from the restricted Hilbert space of the two qubits) and $\mathcal{L}$ is the Liouvillian matrix. Due to the form of the initial quantum state established from Eq.~\eqref{eq:zdfdzdfdzf} and Eq.~\eqref{eq:zdfdzdfdzf222}, which restricts the dynamics of the system to the zero--excitation and one--excitation sectors only, there are just nine nonzero matrix elements of the density vector $\rho$. This dynamical vector can be readily rearranged into the $4\times4$ density matrix
\begin{equation}
\label{eq:dfgdfgdfg}
 \rho =
 \begin{pNiceArray}{cccc}[columns-width=auto]
 \cos^2 \left( \tfrac{\theta}{2} \right) & \langle \sigma_a^\dagger \rangle & \langle \sigma_b^\dagger \rangle & 0 \\
  \langle \sigma_a \rangle & \langle \sigma_a^\dagger \sigma_a \rangle & \langle \sigma_b^\dagger \sigma_a \rangle & 0 \\
    \langle \sigma_b \rangle & \langle \sigma_a^\dagger \sigma_b \rangle & \langle \sigma_b^\dagger \sigma_b \rangle & 0 \\
      0 & 0 & 0 & 0 
\end{pNiceArray},
\end{equation}
which is made explicit using the first and second moments already provided in Sec.~\ref{sec_dynamics}. The knowledge of this two--qubit quantum state, as fully given in Eq.~\eqref{eq:dfgdfgdfg}, allows for the degree of entanglement to be judged using the concurrence of Wootters~\cite{Wootters1998}. Concurrence is defined by the rule $C = \mathrm{max}\{ 0, \lambda_1 - \lambda_2 - \lambda_3 - \lambda_4 \}$, where $\lambda_n$ are the four eigenvalues (in order of decreasing size) of the operator $R$. Using the Pauli matrix $\sigma_y$ and the complex conjugate state $\rho^\ast$, the spin-flipped state $\tilde{\rho}$ for two qubits is given by the transformation $\tilde{\rho} = (\sigma_y \otimes \sigma_y) \rho^\ast (\sigma_y \otimes \sigma_y)$, which then gives access to the operator $R$ via its squared expression $R^2 = \sqrt{\rho} \tilde{\rho} \sqrt{\rho}$. Notably, the concurrence $C$ is zero if the state is unentangled, while it is increasingly positive for quantum states with greater degrees of entanglement. As the quantum state $\rho$ of the two-component quantum battery evolves it is therefore associated with the time-dependent concurrence
\begin{equation}
\label{eq:conccc} 
C_\theta = \sin^2 \left( \tfrac{\theta}{2} \right) \lvert \sin \left( 2Jt \right) \rvert.
 \end{equation}
This dynamical expression is plotted in Fig.~\ref{FIGentangle}~(c) for five selected values of the Bloch sphere polar angle $\theta$. Notably, with $m = 0 , 1, 2, ...$ being a non-negative integer, the entanglement is strongest for a given angle $\theta$ when the dimensionless time $J t = (1+2m)\pi/4$, such that the concurrence becomes $C_\theta = \sin^2 (\theta/2)$. Conversely, there is no charger--battery entanglement ($C_\theta = 0$) at the specific instances in time where $J t = m\pi/2$. Therefore, at the energetic optimal time $t_E = \pi/2J$ for example, the joint quantum state $\rho$ is unentangled since a $\cos (\theta/2) \ket{0}_b + \exp{(\mathrm{i} \phi)} \exp{(-\mathrm{i} \pi \omega_b/2 J)} \sin (\theta/2) \ket{1}_b$ superposition state in the battery qubit has essentially been realized, with the charger qubit lying in its ground state $\ket{0}_a$. Alternatively, at the specific instant $J t = \pi/4$ (and furthermore where the angle $\theta = \pi$, corresponding to the red line in the plot) the global maximum entanglement of $C_\pi = 1$ is seen to be achieved, which ties in with the system establishing a Bell state $(\ket{1}_a \otimes \ket{0}_b -\mathrm{i} \ket{0}_a \otimes \ket{1}_b)/\sqrt{2}$ where the mean populations of the charger and battery qubits are equal. While for this single--cell quantum battery model the impact of quantum entanglement is arguably rather minor, the potentially beneficial effects of entanglement for quantum batteries built with multiple cells will be the subject of a future investigation.


\section{\label{sec_diss}Dissipation}

In any realistic experimental quantum battery system there will be some degree of energy loss into the external environment~\cite{Khodadad2025, DowningPRE}. However, compared to the timescale required for the energy storage phase, the battery qubit should be essentially lossless in order to be practicable as a device. Unfortunately, this inaccessibility suggests that it could be hard to drive the battery qubit directly in the first place, motivating the presence of the charger qubit as a mediator [cf. the setup of Fig.~\ref{figCART}~(b)]. Although for an accessible charger qubit that is easy to prepare in some quantum state, it is reasonable to assume that it can suffer more appreciably from dissipation. In this case, the Liouville–von Neumann equation introduced at the start of Sec.~\ref{sec_dynamics} can be upgraded into the quantum master equation~\cite{Manzano2020}
\begin{equation}
\label{eq:jfhfghfghfg} 
 \partial_t \rho = - \mathrm{i} [ \hat{H}, \rho ] + \tfrac{\gamma}{2} \left( 2 \sigma_a \rho \sigma_a^\dagger - \sigma_a^\dagger \sigma_a \rho - \sigma_a^\dagger \sigma_a \rho \right),
\end{equation}
written in standard Gorini-Kossakowski-Sudarshan-Lindblad form, where $\gamma \ge 0$ is the decay rate from the charger qubit. It is then useful to define an auxiliary frequency $G$ to account for the competition between the coherent coupling rate $J$ and the loss rate $\gamma$ via
\begin{equation}
\label{eq:jfhfghjyjyjyjyjyjyjfghfg} 
G = \sqrt{J^2 -  \left( \tfrac{\gamma}{4} \right)^2 },
\end{equation}
and we will assume the commonly experimentally achievable strong coupling regime where $J > \gamma/4$ in what follows. Using the more sophisticated Eq.~\eqref{eq:jfhfghfghfg} to describe the time evolution of the quantum battery system via the density matrix $\rho$, complete with the Lindbladian dissipator term, leads to some small modifications to the calculations presented in Sec.~\ref{sec_dynamics}. For example, in the dynamical matrix $\mathcal{H}$ of Eq.~\eqref{eq:dfsefsdfsdfdgdfgfsdfs} three diagonal elements should be updated before solving the equation of motion of Eq.~\eqref{eq:dfsdfs}, namely $\mathcal{H}_{11} \to \mathcal{H}_{11} - \mathrm{i}\gamma$, $\mathcal{H}_{33} \to \mathcal{H}_{33} - \mathrm{i}\gamma/2$ and $\mathcal{H}_{44} \to \mathcal{H}_{44} - \mathrm{i}\gamma/2$. Hence, when dissipation is properly included into the quantum model, the dynamical stored energy $E$ of Eq.~\eqref{eq:sfddddddffsdfsf} is superseded by
\begin{equation}
\label{eq:jfhfghfgbgngngngjhfg} 
E_\theta = \omega_b \sin^2 \left( \tfrac{\theta}{2} \right) \left( \tfrac{J}{G} \right)^2 \sin^2 \left( G t \right) \mathrm{e}^{-\frac{\gamma t}{2}},
\end{equation}
which features both a characteristic exponential decay (with the time constant $2/\gamma$) and a renormalization of the effective Rabi frequency from $J$ to the $G$ of Eq.~\eqref{eq:jfhfghjyjyjyjyjyjyjfghfg}. Consequentially, the optimal charging time $t_E$ of Eq.~\eqref{eq:dvddeddfg} and the maximum stored energy $E(t_E)$ of Eq.~\eqref{eq:dfgdgdfgbv} should be replaced by the exact formulae
\begin{align}
\label{eq:dfgdfghhg} 
t_E &= \tfrac{ \arctan{ \left( \tfrac{4G}{\gamma} \right) }  }{G},
\\
E_\theta (t_E) &= \omega_b \sin^2 \left( \tfrac{\theta}{2} \right) \mathrm{e}^{-\frac{\gamma}{2G} \arctan{ \left( \frac{4G}{\gamma} \right) }}.
\end{align}
Carrying out series expansions of these optimal energetic expressions makes their deviations from the dissipationless case more explicit, and deep into the strong coupling regime, where $J \gg \gamma$ holds true, leads to the leading order corrections
\begin{align}
\label{eq:dfgdfbgngfgnghhg} 
t_E &\simeq \tfrac{\pi}{2J} \left( 1 - \tfrac{\gamma}{2\pi J} \right),
\\
E_\theta (t_E) &\simeq \omega_b \sin^2 \left( \tfrac{\theta}{2} \right) \left( 1 - \tfrac{\pi \gamma}{4 J} \right).
\end{align}
Clearly while the presence of a lossy charger qubit reduces the optimal charging time $t_E$, it also degrades the quantity of energy $E(t_E)$ stockpiled in the battery qubit. Both modifications are relatively minor, with the changes being by a certain amount linear in the small dimensionless parameter $\gamma/J$.

The optimal performance of the quantum battery with regard to its charging power is similarly slightly reconstructed from the lossless expressions of Eq.~\eqref{eq:dvdxfgxffdvdsgfg} and Eq.~\eqref{eq:dfxgfxgfggdg}. The exact charging time $t_P$ can be found from the generalized transcendental equation $( 1+ \gamma x / 2 G) \tan(x) = 2 x$, written in the dimensionless variable $x = G t_P$. Let us suppose a perturbative solution in the form $x = A - K \gamma/G$, where the number $A$ is defined in Eq.~\eqref{eq:dvdxfgxffdvdsgfg} such that the proposed trial solution is exact in the dissipationless limit, and where $K$ is a constant to be found. Then, working well inside of the strong coupling limit where $J \gg \gamma$, one finds that the number $K = - A \tan(2A)/4 = 0.306\ldots$ such that the optimal charging time $t_P = (A - K \gamma/G)/G$. Therefore, the leading order corrections to the optimal charging time and power are given by
\begin{align}
\label{eq:dfgdfghvcvcvcvhg} 
t_P &\simeq \tfrac{A}{J} \left( 1 - L \tfrac{\gamma}{J} \right),
\quad\quad L = 0.262 \dots,
\\
P_\theta (t_P) &\simeq B \omega_b J \sin^2 \left( \tfrac{\theta}{2} \right) \left( 1 - \tfrac{A}{2} \tfrac{\gamma}{J} \right),
\end{align}
where the introduced numerical coefficient $L = K/A$. Akin to the behaviour for the stored energy, the optimal charging time $t_P$ is arrived at sooner due to the damping process accounted for in Eq.~\eqref{eq:jfhfghfghfg}, while the peak power $P (t_P)$ is intuitively reduced by a small amount due to the energy loss.

The general expression found for the ergotropy $\mathcal{E}$, as presented in Eq.~\eqref{eq:fsdfs}, also changes when dissipation is considered. In a similar manner as to how the effective Hamiltonian $\mathcal{H}$ needed to be updated when determining the second moments [see the discussion above Eq.~\eqref{eq:jfhfghfgbgngngngjhfg}], the dynamical matrix $\mathcal{M}$ of Eq.~\eqref{eq:dfsefsdfsdfsdfs} should see three elements replaced following the rules $\mathcal{M}_{11} \to \mathcal{M}_{11} -\mathrm{i}\gamma/2$, $\mathcal{M}_{33} \to \mathcal{M}_{33} -\mathrm{i}\gamma$ and $\mathcal{M}_{44} \to \mathcal{M}_{44} -\mathrm{i}\gamma/2$ before calculating the odd moments of the system from the equation of motion of Eq.~\eqref{eq:sdfsfdsdfdsf}. The resulting ergotropy $\mathcal{E}$ can then be neatly given in terms of the useful dynamical function $f(t)$ as
\begin{widetext}
\begin{equation}
\label{eq:jfhfgdfsdfhfghfg} 
\mathcal{E}_\theta = \omega_b \left[ \sin^2 \left( \tfrac{\theta}{2} \right) f^2(t) - \tfrac{1}{2} \left( 1 - \sqrt{1 - 4 \sin^4 \left( \tfrac{\theta}{2} \right) f^2(t) \left[ 1 - f^2(t) \right] } \right)  \right],
\quad\quad\quad\quad
f(t) =  \tfrac{J}{G}  \sin \left( G t \right) \mathrm{e}^{-\frac{\gamma t}{4}}.
\end{equation}
\end{widetext}
This efficacious analytical result fully describes the work which can be extracted from the battery qubit after being charged by the mediating charger qubit (which is itself prepared in an arbitrary state accounted for by the Bloch angle $\theta$, and despite the presence of dissipation occurring at the rate $\gamma$). Similarly, using the ubiquitous dynamical function $f(t)$, the battery capacity $\mathcal{K}$ generalizes to the compact expression [cf. Eq.~\eqref{eq:fghjhgkuguk}]
\begin{equation}
\label{eq:codfgdfgdnccc} 
\mathcal{K}_\theta = \omega_b \sqrt{ 1 - 4 \sin^4 \left( \tfrac{\theta}{2} \right) f^2(t) \left[ 1- f^2(t) \right] }, 
 \end{equation}
 which captures the necessary degradation of the battery population due to dissipation from the charger. We hope that the brief open quantum systems analysis of this section has some utility for describing experiments with quantum batteries built from coupled two-level systems, and furthermore where energy loss is unfortunately non-negligible.


\section{\label{sec_disc}Discussion}

We have revisited the problem of charging a quantum battery composed of a pair of coupled qubits, with the natural generalization of allowing for the initial quantum state of the charger qubit to lie anywhere on the surface of the Bloch sphere. The amount of energy stored in the battery can be described straightforwardly [cf. Eq.~\eqref{eq:sfddddddffsdfsf}] in a manner reminiscent of the quantum area theorem~\cite{Allen1987, Shore2011}. We have gone on to derive simple analytical expressions describing the exact ergotropy and battery capacity of this device given an arbitrary starting charging state [cf. Eq.~\eqref{eq:fsdfs} and Eq.~\eqref{eq:fghjhgkuguk}]. These thermodynamic quantities can indeed take on desirable nonzero values, which can be ascribed to population inversions or quantum coherences (or a mixture of both mechanisms) depending upon the value of the starting Bloch polar angle [cf. Eq.~\eqref{eq:dgffgfg} and Eq.~\eqref{eq:dgffgfg22222}]. We have also shown how the ergotropic charging power deviates from the standard charging power in a meaningful way [cf. the final column of Fig.~\ref{FIGthermo}]. Conveniently, the thermodynamically optimal charging protocol can be captured by two simple approximate expressions for the best charging time and for the associated peak charging power [cf. Eq.~\eqref{eq:dgsfsdf} and Eq.~\eqref{eq:sdfsdf}]. Finally, we have studied the leading order effects of dissipation throughout Sec.~\ref{sec_diss}, which completes our comprehensive analysis of this fundamental quantum thermodynamic system.  

We hope that our basic quantum theory can guide future experiments employing coupled qubits as quantum batteries, especially those carried out in a similar vein to the insightful experiments of Ref.~\cite{Niu2024}. Our groundwork calculations, on what is arguably a fundamental building block for a whole class of quantum battery designs, may also act as a springboard for more sophisticated many--body theoretical investigations within the ever expanding realm of quantum energy science~\cite{Zhang2025Zhang2025, Siddique2025}.
\\


\noindent \textbf{Acknowledgments}\\
\textit{Funding}: CAD is supported by the Royal Society via a University Research Fellowship (URF\slash R1\slash 201158). \textit{Discussions}: CAD thanks André Cidrim (Universidade Federal de São Carlos) and Alan Costa dos Santos (Instituto de Física Fundamental -- CSIC) for discussions as part of their Royal Society International Exchanges grants (IES/R1/241078) and (IES/R1/251033) respectively, as well as Aiman Khan (University of Portsmouth) for conversations held at an early stage of this research. \textit{Data and materials availability}: All data is available in the manuscript.
\\



\end{document}